\documentclass[11pt,bibnotes,aps,prb,twocolumn,reprint]{revtex4-1}
\usepackage{graphicx}
\usepackage{amsmath}

\begin{document}
\title{Hydrodynamic electron transport and nonlinear waves in graphene}
\author{D. Svintsov}
\affiliation
{Institute of Physics and Technology, Russian Academy of Science, Moscow 117218}
\affiliation
{Moscow Institute of Physics and Technology, Dolgoprudny 141700, Russia}

\author{V. Vyurkov}
\affiliation
{Institute of Physics and Technology, Russian Academy of Science, Moscow 117218}
\affiliation
{Moscow Institute of Physics and Technology, Dolgoprudny 141700, Russia}

\author{V. Ryzhii}
\affiliation 
{Research Institute for Electrical Communication, Tohoku University, Sendai 980-8577,  Japan}

\author{T. Otsuji}
\affiliation
{Research Institute for Electrical Communication, Tohoku University, Sendai 980-8577,  Japan} 

\begin{abstract}
We derive the system of hydrodynamic equations governing the collective motion of massless fermions in graphene. The obtained equations demonstrate the lack of Galilean and Lorentz invariance, and contain a variety of nonlinear terms due to quasi-relativistic nature of carriers. Using those equations, we show the possibility of soliton formation in electron plasma of gated graphene. The quasi-relativistic effects set an upper limit for soliton amplitude, which marks graphene out of conventional semiconductors. The mentioned non-invariance of  equations is revealed in spectra of plasma waves in the presence of steady flow, which no longer obey the Doppler shift. The feasibility of plasma wave excitation by direct current in graphene channels is also discussed.

\end{abstract}

\maketitle

\section{Introduction}
The models of carrier transport in graphene should account for strong carrier-carrier interaction~\cite{Electron-interactions,Interactions-transport}, which is governed by a large ''fine structure constant'' $e^2/(\hbar v_F)\sim 1$ and logarithmically divergent collision integral for collinear scattering~\cite{Kashuba, Quantum-critical}. The relative strength of carrier-carrier scattering compared to other relaxation mechanisms was also proved in the transparency measurements of optically pumped graphene~\cite{Transparency-1,Transparency-2}. The corresponding relaxation time is estimated to be less than 100~fs.

The most natural way to account for carrier-carrier interactions in transport models is to use local equilibrium (hydrodynamic) distribution functions as a first approximation to the solution of kinetic equation\cite{Ford-Uhlenbeck}. Several approaches for description of hydrodynamic transport in graphene were presented in Refs.~\cite{Our-hydrodynamics,Polini-hydro,MacDonald-HD, A-nearly-perfect-fluid,  Mendoza, Quantum-critical-magnetotransport,Collisionless-HD-magnetic, Schutt-drag}. Within hydrodynamic models, it is possible to explain the temperature-independent dc conductivity of graphene in charge neutrality point~\cite{Our-hydrodynamics}, and the strong coulomb drag between electrons and holes~\cite{Our-hydrodynamics, Levitov-drag,Schutt-drag}. The other predictions of hydrodynamic transport, such as pre-turbulent current flow due to low viscosity~\cite{A-nearly-perfect-fluid}, existence of electron-hole sound~\cite{Our-hydrodynamics}, and current saturation at high electric fields due to heating of electrons~\cite{MacDonald-HD}, still expect their experimental verification.

In recent works on graphene hydrodynamics~\cite{A-nearly-perfect-fluid,Polini-hydro,Collisionless-HD-magnetic}, the equations were obtained under assumption of low drift velocity $u$ of electron plasma ($u \ll v_F$). Several nonlinear terms were inevitably lost under such assumption. In several other works~\cite{Quantum-critical-magnetotransport,Mendoza}, the hydrodynamics of massless quasiparticles in graphene was obtained from hydrodynamics of ultrarelativistic plasma by a simple replacement of the speed of light $c$ by the Fermi velocity $v_F$. Such a spurious analogy is misleading in this particular case as electrons in graphene are neither Galilean-~\cite{Non-gallilean}, nor truly Lorentz-invariant system (in general, this refers to any electrons in solids). The reason is that for velocities less and of the order of $v_F\simeq c/300$, the distortion of space-time metrics is negligible. Hence, dealing with quasi-relativistic particles in Galilean space-time, one will obtain hydrodynamic equations that are neither Galilean, nor Lorentz-invariant.

In this paper, we present a rigorous derivation of hydrodynamic equations for massless electrons in graphene, following the general strategy put forward by Achiezer {\textit{et.~al.}}~\cite{Akhiezer}. Besides, we eliminate the restriction on the flow velocity to be much less than the Fermi velocity. This opens up an opportunity to study a wide variety of nonlinear phenomena, such as propagation of large amplitude waves~\cite{Shock-and-solitonlike-waves,Rudin-Samsonidze}, photovoltaic response~\cite{Graphene-photovoltaic}, transport at high current flows~\cite{Large-current}, and acousto-electronic interactions~\cite{Acousto-electronics}.

In graphene it is impossible to introduce a constant electron effective mass $m$ as a proportionality coefficient binding the momentum to the velocity. Consequently, the Euler equation can be no longer presented in canonical form  $\partial_t \mathbf{u} + (\mathbf{u} \nabla) \mathbf{u} + (\nabla P)/\rho = 0$,
where $\mathbf{u}$ is the drift velocity, $\rho$ is the mass density of electrons, and $P$ is the pressure. However, a fictitious (hydrodynamic) mass ${\cal M}$ depending on the particle density $n$ and the drift velocity naturally arises in the Euler equation. One more unusual feature of Euler equation is an appearance of density- and velocity dependent factor before the convection term $(\mathbf{u} \nabla) \mathbf{u} $. We show that in the degenerate electron system this term vanishes, giving way to a weaker fourth-order nonlinearity proportional to $u^2 (\mathbf{u} \nabla) \mathbf{u} $.

Using the derived equations, we study the nonlinear effects in plasma wave propagation in graphene. Hydrodynamics proved to be an extremely efficient tool for the study of electron plasma in two-dimensional (2D) electron systems~\cite{Chaplic-Old,Chaplic-solitons,Dyakonov-Shur,Dyakonov-Shur-Choking}. Collective dynamics of electrons in two dimensions has a rich analogy with the hydrodynamics of liquids, including the phenomena of electron flow choking~\cite{Dyakonov-Shur-Choking} and formation of shallow- and deep-water plasma waves in gated and non-gated systems, respectively~\cite{Dyakonov-Shur}. Despite huge efforts in the field of graphene plasmonics~\cite{Graphene-plasmonics,Plasmonics-for-tunable-THz,Plasmons-imaging, Plasmons-RPA, Ryzhii-plasma}, the problem of nonlinear plasma waves stayed beyond the scope of recent works.

We show that the balance between nonlinearities and dispersion allows the formation of solitary plasma waves in gated graphene. We find that 'relativistic' terms in Euler equation set an upper limit for the soliton amplitude and broaden its profile. This much differs from the solitons in systems of massive 2D electrons~\cite{Chaplic-solitons,Solitons-2DEG}, which behave similarly to the solitary waves in water.

The features of electron hydrodynamics are also pronouncedly revealed in the spectra of collective (plasma) excitations in graphene in the presence of stationary electron flow with velocity $u_0$. It is natural to expect that velocities of forward and backward plasma waves are $u_0 \pm s_0$, where $s_0$ is the wave velocity in electron fluid at rest. We demonstrate, however, that in graphene the dependence of the wave velocities on the flow velocity is more complicated due to the lack of Galilean and Lorentz invariance. In particular case of strongly degenerate electron system, these velocities are $\frac{1}{2}u_0 \pm s_0$.

We also revealed the potentiality of plasma instability in gated graphene in the presence of steady current. We prove that this effect (predicted for high-mobility 2D electron systems based on the conventional semiconductors by Dyakonov and Shur~\cite{Dyakonov-Shur}) persists for graphene with its unusual hydrodynamics. We find the ultimate increment of plasma waves and show that this instability could be realized in graphene channels of submicron length.  

The work is organized as follows. In section II we derive the set of hydrodynamic equations and discuss the terms arising due to the massless nature of Dirac fermions. In section III we demonstrate several solutions of those equations revealing the features of electrons in graphene. Particularly, we obtain the profiles of solitary waves in gated graphene, find the spectra of plasma waves in the presence of steady flow, and show the possibility of plasma wave self-excitation under certain boundary conditions. The main results are discussed in section IV. Some mathematical details concerning the derivation of equations are singled out into Appendix. 

\section{Derivation of hydrodynamic equations}
We consider 2D plasma of massless electrons in graphene with a linear dispersion law $\epsilon_{\bf p} = p v_F$. We assume the Fermi level to be above the Dirac point and neglect the contribution of holes. The effects of electron-hole interaction on transport in graphene were extensively studied in Refs.~\cite{Quantum-critical, Kashuba, Our-hydrodynamics, Quantum-critical-magnetotransport}.

The starting point for derivation of hydrodynamic equations lies in the construction of the distribution function of carriers which turns the collision integral to zero. Regardless of the energy spectrum $\epsilon_{\bf p}$ this function is
\begin{equation}
\label{F_e}
f({\bf p})=\left[1+\exp\left(\frac{\epsilon_{\bf p}-{\bf p u}-\mu}{T}\right)\right]^{-1},
\end{equation}
where the quantities defined from the hydrodynamic equations are the chemical potential $\mu$, the drift velocity ${\bf u}$, and the temperature $T$ (measured in the energy units).

The set of hydrodynamic equations is obtained by integrating the kinetic equation timed by $1$, $p_i$, and $\epsilon_{\bf p}$ over the phase space. The kinetic equation for massless electrons reads
\begin{equation}
\frac{\partial f}{\partial t} + v_F\frac{\bf p}{p}\frac{\partial f}{\partial {\bf r}} + {\bf F}\frac{\partial f}{\partial {\bf p}} = St_{e-i}\{ f \}+St_{e-e}\{ f \}.
\end{equation}
Here $St_{e-i}\{ f \}$ includes the electron-impurity  and electron-phonon collision integrals, $St_{e-e}\{ f \}$ is the electron-electron collision integral, ${\bf F} = e\partial\varphi/\partial {\bf r}$ is the force acting on electron, $e = |e|$. The dissipative terms in hydrodynamic equations due to the electron-impurity and electron-phonon collisions were discussed in previous works~\cite{Our-hydrodynamics,MacDonald-HD}, and further will be omitted. 

It is instructive that all statistical average values like the electron density $n = 4 \sum_{\bf p} f_{\bf p}$, flux ${\bf j} = 4 \sum_{\bf p}{\bf v}_{\bf p} f_{\bf p}$, and internal energy density $\varepsilon = 4 \sum_{\bf p}\epsilon_{\bf p} f_{\bf p}$ can be calculated exactly with the distribution function~(\ref{F_e}) when the spectrum $\epsilon_{\bf p}$ is linear (see the derivation in Appendix), namely,
\begin{gather}
\label{Eqn-of-state}
n = \frac{n_0}{\left[1- u^2 / v_F^2\right]^{3/2}},\\
{\bf j} = n \bf{u},\\
\varepsilon = \frac{\varepsilon_0}{\left[1 - u^2/v_F^2 \right]^{5/2}}.
\end{gather}
Here $n_0$ and $\varepsilon_0$ are the steady-state particle density and energy density, respectively, given by
\begin{gather}
n_0 = \frac{2 T^2}{\pi \hbar^2 v_F^2} \int\limits_0^\infty \frac{t dt}{1+e^{t-\mu/T}},\\
\varepsilon_0 =  \frac{2 T^3}{\pi \hbar^2 v_F^2} \int\limits_0^\infty \frac{t^2 dt}{1+e^{t-\mu/T}}.
\end{gather}

For brevity, we introduce the pressure of electron plasma
\begin{equation}
\label{Pressure}
P=\frac{ \varepsilon }{2},
\end{equation}
and an analogue of the electron mass density
\begin{equation}
\label{Mass_density}
\rho =\frac{3 \varepsilon }{2v_{F}^{2}}.
\end{equation}

In these notations, the set of hydrodynamic equations takes on the form:
\begin{equation}
\label{Continuity}
\frac{\partial n }{\partial t}+\frac{\partial (n u_i)}{\partial x_i} = 0,
\end{equation}
\begin{equation}
\label{Euler}
\frac{\partial  \left( \rho  u_i \right)}{\partial t} + \frac{\partial \Pi_{ij}}{\partial x_j} - e n \frac{\partial \varphi }{\partial x_i}=0.
\end{equation}
Equations (\ref{Continuity}) and (\ref{Euler}) represent the continuity and the Euler equation, respectively. The elements of stress tensor are (the velocity ${\bf u}$ is directed along $x$-axis)
\begin{gather}
\Pi_{xx} = P \left[ 1 + 2(u/v_F)^2\right],\\
\Pi_{yy} = P \left[ 1 - 2(u/v_F)^2\right].
\end{gather}
The heat transfer equation (which will not be discussed here in detail) reads
\begin{equation}
\frac{\partial \varepsilon}{\partial t} + v_F^2\frac{\partial (\rho u_i)}{\partial x_i} + ({\bf F},{\bf j}) = 0.
\end{equation}

It is important that the electron sheet density $n$ arises in the continuity equation, while the mass density $\rho$ arises in the Euler equation. Those quantities are not directly proportional to each other as it is impossible to introduce a constant electron effective mass $m$. In other words, the fictitious particle mass $ {\cal M} = \rho / n$ is density- and velocity-dependent and cannot be factored out of the differential operator. In the degenerate electron system ($\mu \gg T$) the expression for mass reads
\begin{equation}
{\cal M} = \frac{\mu/v_F^2}{\sqrt{1 - u^2/v_F^2}}.
\end{equation}

To recognize the peculiarities of the obtained Euler equation and analyze the emerging nonlinearities, it would be convenient to present it in the canonical form (further we restrict ourselves with one-dimensional motion). Excluding the time derivatives of density $\rho$ with the use of Eq. (\ref{Eqn-of-state}) [see also Eq.~(\ref{Derivatives})], we arrive at the following equation: 
\begin{multline}
\label{Canonical-graphene}
\frac{\partial u}{\partial t} \left[ 1+\frac{\beta^2 \left( 5-6\xi  \right)}{1-\beta^2}\right]+u \frac{\partial u}{\partial x} \left[ \left( 3-4\xi  \right)-\frac{\beta^2 \left( 5-6\xi  \right)}{1-\beta^2} \right]+\\
+\frac{2 \xi v_F^2}{3 n}\frac{\partial n}{\partial x} \left( 1- \beta^2 \right)-\frac{n}{\rho} \frac{\partial (e \varphi)}{\partial x}  = 0.
\end{multline}
Here we have introduced the relativistic factor $\beta = u/v_F$ and the dimensionless function $\xi$ characterizing the thermodynamic state of the electron system:
\begin{equation}
\xi =\frac{n^2}{\varepsilon \left\langle {\varepsilon^{-1}} \right\rangle},
\end{equation}
where $\langle \varepsilon^{-1} \rangle \neq \varepsilon^{-1}$ is the density of inverse energy [see Eq.~(\ref{Inverse-energy})]. The function $\xi$ varies from $1/2$ at $\mu/T\to - \infty$ to $3/4$ at  $\mu/T\to + \infty$. At $\mu \gg T$ it is given by the following asymptotic relation 
\begin{equation}
\xi_{\mu\gg T} = \frac{3}{4}\left( 1 - 2\frac{n_T}{n}\right).
\end{equation}
Here $n_T$ is the density of thermally activated electrons at $\mu=0$.

In hydrodynamic equations for massive particles, two sources of nonlinearities exist: the current density term in continuity equation $\partial_x(n u)$, and the nonlinear convection term in Euler equation $u\partial_x u$. Much greater variety of nonlinearities is involved in the Euler equation for electrons in graphene (\ref{Canonical-graphene}). They can be classified as 'relativistic' nonlinearities due to high drift velocities and nonlinearities due to density dependence of hydrodynamic mass ${\cal M}$. To compare their 'strength', we consider small perturbations of density, velocity, and electric potential: $n = n_0 + \delta n(x,t)$, $u = \delta u(x,t)$, $\varphi = \varphi_0 + \delta \varphi(x,t)$.

It is easy to see that the 'relativistic' nonlinearities are, at least, the third-order terms. Dropping them, we can rewrite the Euler equation as
\begin{equation}
\frac{\partial u}{\partial t} + u \frac{\partial u}{\partial x} \left( 3-4\xi  \right)+\frac{1}{{\cal M}}\left[ \frac{1}{n}\frac{\partial P}{\partial x} - \frac{\partial (e\varphi)}{\partial x}\right] = 0.
\end{equation}
The nonlinear convection term in Euler equation is weakened due to the factor $3-4\xi < 1$. In degenerate electron system this term can be estimated as
\begin{equation}
6\frac{n_T}{n_0}\delta u \frac{\partial \delta u}{\delta x} \approx 6\frac{n_T}{n_0} \frac{s_0^2}{n_0^2} \delta n \frac{\partial \delta n}{\delta x}.
\end{equation}
Here $s_0$ is the velocity of collective excitations (plasma waves) in graphene. At low temperatures and elevated Fermi energies this term becomes infinitesimal and surrenders to the higher-order nonlinearity $u^3\partial_x u$.

In comparison with the convective term, the nonlinearity due to the density dependent mass (${\cal M}\propto n^{1/2}$) is much stronger. The corresponding term could be evaluated as
\begin{equation}
\label{Mass-contribution}
-\frac{1}{2{\cal M}}\frac{\delta n}{n_0}\delta \left[ \frac{1}{n_0}\frac{\partial P}{\partial x} - \frac{\partial (e\varphi)}{\partial x}\right] \approx - \frac{s_0^2}{2 n_0^2}\delta n\frac{\partial \delta n}{\partial x}.
\end{equation} 

It is readily seen from the above considerations that nonlinear transport phenomena in graphene are rather governed by the density dependence of mass ${\cal{M}}$ than by nonlinear convection, which used to occur in common semiconductors.

\section{Nonlinear effects in plasma wave propagation}

\subsection{Formation of solitons in gated graphene}
In gated structures the electron density $n$ is related to the local electric field $-\partial\varphi/\partial x$ in graphene via a weak nonlocality approximation
\begin{equation}
\label{Gradual_channel}
\frac{\partial (e\varphi)}{\partial x} = -\frac{4\pi e^2 d_1 d_2}{\kappa (d_1+d_2)}\frac{\partial n}{\partial x} - \frac{4\pi e^2 d_1^2 d_2^2}{3\kappa (d_1+d_2)}\frac{\partial^3 n}{\partial x^3}.
\end{equation}
Here $d_1$ and $d_2$ are distances from the graphene layer to the top and bottom gates, respectively, and $\kappa$ is the gate dielectric permittivity. The third derivative term describes a weak dispersion of plasma waves in gated structures. A subtle balance between dispersion and nonlinearity results in the formation of solitary waves.

We search for the solutions of hydrodynamic equations in the form $n = n_0 + \delta n(z)$, $u = \delta u(z)$, where $z = x - u_0 t$ is the running coordinate, and $u_0$ is the soliton velocity being slightly different from the plasma wave velocity $s_0$ due to the dispersion. Within the hydrodynamic model, the expression for plasma wave velocity can be represented as~\cite{Our-hydrodynamics}
\begin{equation}
s_0^2 = v_F^2 \frac{2\xi}{3}\left(1 + \frac{4\pi e^2}{\kappa} \frac{d_1 d_2}{d_1+d_2} \langle\varepsilon^{-1}\rangle \right).
\end{equation}

Integration of the continuity equation (\ref{Continuity}) provides the relation between $u$ and $n$:
\begin{equation}
\label{Velocity-in-soliton}
u = u_0\frac{n-n_0}{n}.
\end{equation}

Eliminating the drift velocity $u$ and the electric potential $\varphi$ from the Euler equation (\ref{Canonical-graphene}) with the help of Eqs.~(\ref{Gradual_channel}, \ref{Velocity-in-soliton}), we arrive at the dynamic equation for solitary wave. The latter is concisely represented using the dimensionless variables $\zeta = z \sqrt{3/(d_1 d_2)}$, $\nu = \delta n/n_0$,  $\beta = u/v_F$, $\beta_0 = u_0/v_F$, $\tilde{s}_0 = s_0/v_F$ as follows:
\begin{equation}
\label{KdV-nonlinear}
F(\nu) \frac{\partial \nu}{\partial \zeta} + \left(\tilde{s}_0^2 - \frac{1}{2}\right)\frac{\partial^3 \nu}{\partial \zeta^3} =0,
\end{equation}
\begin{multline}
F(\nu) = \tilde{s}_0^2 - \frac{1}{2} + \frac{\beta \beta_0^2 \left(\beta+\beta_0 \right)}{2 (\nu +1)^{3/2} \left(1-\beta^2\right)}+\\
\frac{1-\beta^2}{2\sqrt{\nu +1}}-\frac{\beta_0^2}{(\nu +1)^{3/2}}.
\end{multline}

On expanding $F(\nu)$ in series over $\nu$, one arrives at the well-known Korteweg-de Vries~\cite{KdV-original} equation
\begin{multline}
\label{Kortveg}
\left(\tilde{s}_0^2 - \beta_0^2 \right)\frac{\partial \nu}{\partial \zeta} + \left(\tilde{s}_0^2 - \frac{1}{2}\right)\frac{\partial^3 \nu}{\partial \zeta^3} + \\
\left(\frac{3}{2}\beta_0^2 - \frac{1}{4}\right)\nu \frac{\partial \nu}{\partial \zeta} = 0.
\end{multline}
The solutions of this equation correspond to the so called bright solitons; their shape is given by~\cite{Kinetics}
\begin{equation}
\label{Soliton-solution}
\delta n(z) = \delta n_{\max} \cosh^{-2}\left[\frac{z}{2}\sqrt{\frac{3}{d_1 d_2} \frac{s_0^2}{2s_0^2 - v_F^2}\frac{\delta n_{\max}}{n_0}} \right].
\end{equation}
The maximum soliton height $\delta n_{\max}$ is bound to its velocity $u_0$ via
\begin{equation}
\delta n_{\max} = \frac{n_0}{2} \frac{u_0^2 - s_0^2}{s_0^2}.
\end{equation}
The soliton width $W$ is
\begin{equation}
W = 2\sqrt{\frac{d_1 d_2}{3}\frac{2s_0^2 - v_F^2}{s_0^2} \frac{\delta n_{\max}}{n_0}}.
\end{equation} 
As $\delta n_{\max} \ll n_0$, and $s_0^2 > v_F^2/2$, the soliton width can be much greater than the distance to the gates, which justifies the applicability of weak nonlocality approximation. Besides, the soliton width should markedly exceed the inelastic (electron-electron) free path to justify the validity of hydrodynamic approach. The free path is less than 100 nm, which follows from the experimental~\cite{Transparency-1,Transparency-2} and theoretical~\cite{Our-hydrodynamics} estimates of collision frequencies.

Apart from the numerical coefficients, the obtained parameters of solitons coincide with those in 2D plasma of massive electrons in Ref.~\cite{Chaplic-solitons}. To reveal the unique features of graphene electron hydrodynamics, one should go beyond the condition $\delta n \ll n_0$ and analyze the general expression for $F(\nu)$. The necessity of rigorous treatment arises when the velocity $u$ approaches the Fermi velocity, i.e. already at $\delta n_{\max}/n_0 \approx v_F/s_0$.

\begin{figure}[ht]
\center{\includegraphics[width=1.0\linewidth]{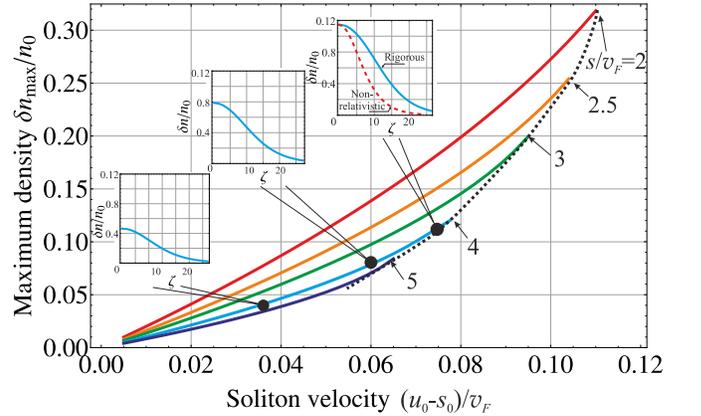} }
\caption{Dependence of soliton amplitude $\delta n_{\max}/n_0$ on its velocity $(u_0 - s_0)/v_F$ at different velocities of plasma waves $s_0$. The dashed black line indicates the boundary of soliton existence. The profiles of solitary waves obtained from numerical integration of Eq.~(\ref{KdV-nonlinear}) are plotted in insets}
\label{Fig-soliton}
\end{figure}

The numerical solution of Eq.~(\ref{KdV-nonlinear}) shows that solitons exist when the maximum particle density $\delta n_{\max}$ lies below a certain critical density. The higher is the plasma wave velocity $s_0$, the lower is the critical density. The relation between $\delta n_{\max}$ and soliton velocity $u_0 - s_0$ is plotted in Fig.~\ref{Fig-soliton}, the termination of the curves corresponds to the critical density and velocity. It is seen from Eq.~(\ref{Soliton-solution}) that the soliton width shrinks as its amplitude increases. The numerical results of solving the rigorous KdV equation (\ref{KdV-nonlinear}) plotted inside the insets in Fig.~\ref{Fig-soliton} indicate that the width of soliton decreases only slightly as its amplitude grows. Given the value of $\delta n_{\max}$, the profile of real soliton is broder that that obtained from 'non-relativistic' approximation (\ref{Kortveg}).

\subsection{'Shallow-water' plasma waves in the presence of steady electron flow}

To obtain the spectra of plasma waves in the presence of steady flow with velocity $u_0$ we linearize the hydrodynamic equations assuming a harmonic time dependence of perturbations
\begin{gather}
n= n_0 + \delta n(x) e^{-i\omega t},\\
u = u_0 + \delta u(x) e^{-i\omega t}.
\end{gather}
This procedure leads to the 'equations of motion' for plasma oscillations
\begin{equation}
\label{Eqn-of-continuity}
-i \omega\delta n + n_0 \frac{\partial \delta u}{\partial x} + u_0 \frac{\partial \delta n}{\partial x} =0,
\end{equation}
\begin{multline}
\label{Eqn-of-motion}
-i \omega \left[1+ \gamma \right] \delta n + u_0 \frac{\partial \delta u}{\partial x} \left[3-4\xi_0 - \gamma \right] + \\
\frac{s_0^2}{n_0} \frac{\partial \delta n}{\partial x} = 0,
\end{multline}
where we have introduced another 'relativistic factor' $\gamma$
\begin{equation}
\gamma = \frac{\beta_0^2}{1-\beta_0^2} (5-6\xi_0).
\end{equation}
A weak dispersion of plasma waves was neglected here.

\begin{figure}[ht]
\begin{minipage}[ht]{0.9\linewidth}
\center{\includegraphics[width=1.0\linewidth]{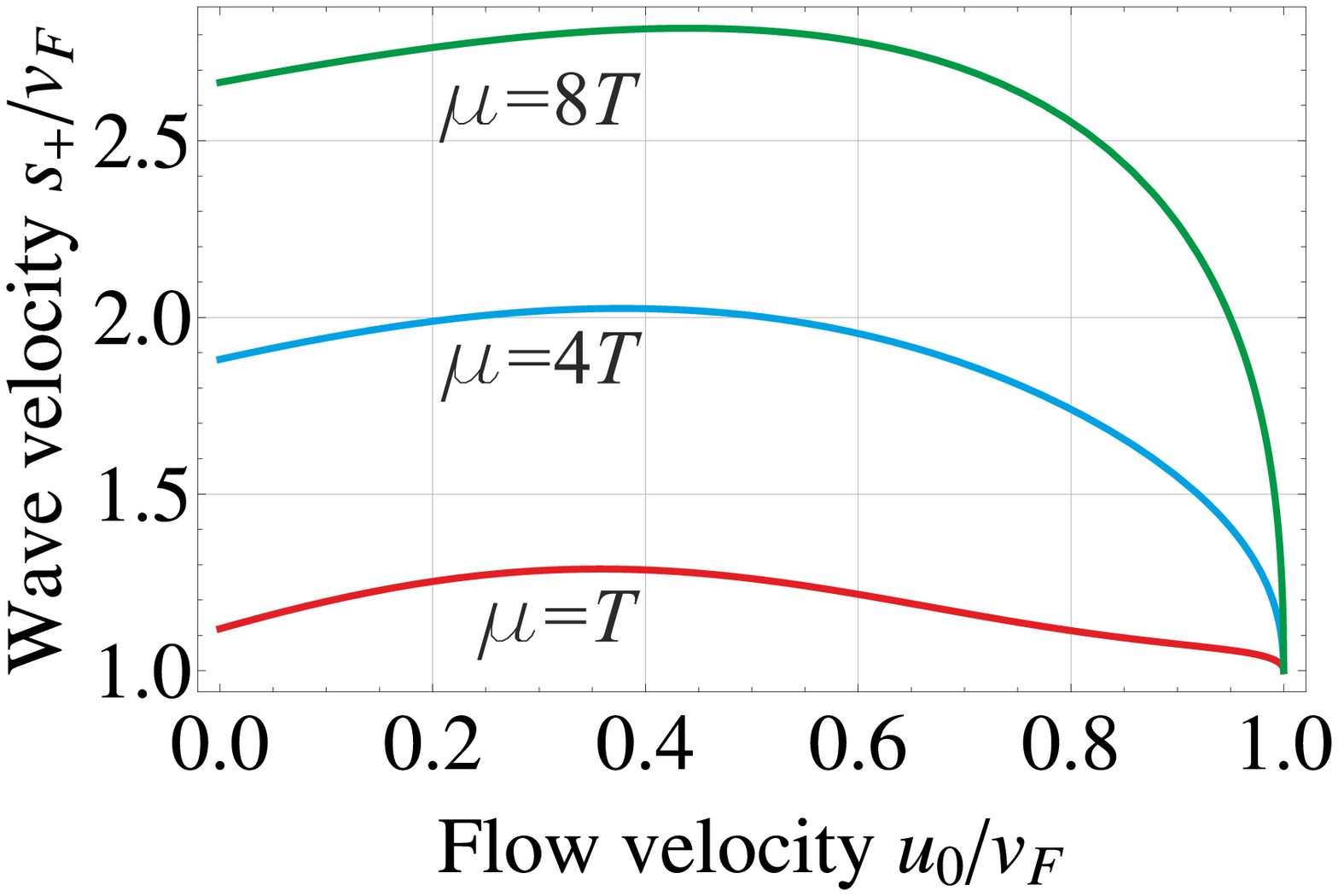} }
\end{minipage}
\vfill
\begin{minipage}[ht]{0.9\linewidth}
\center{\includegraphics[width=1.0\linewidth]{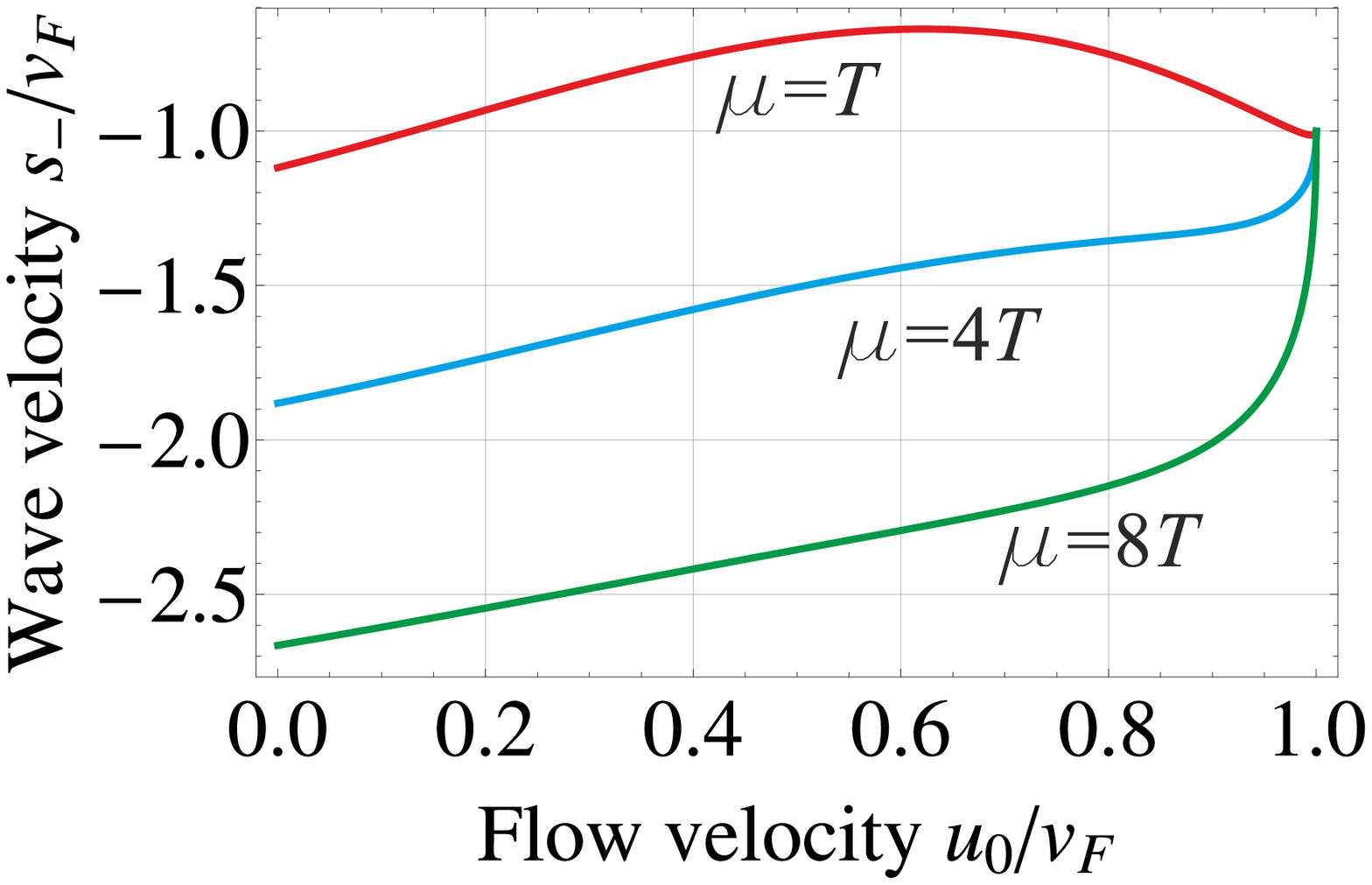} }
\end{minipage}
\caption{Dependencies of plasma wave velocities $s_\pm$ in the presence of steady flow on the flow velocity $u_0$}
\label{Spm}
\end{figure}

Assuming a harmonic dependence of all quantities on the coordinate, that is $\propto e^{i k x}$, we obtain the linear law of plasma wave dispersion $\omega_\pm = s_\pm k$. The velocities of forward ($s_+$) and backward ($s_-$) waves are given by 
\begin{equation}
\label{Disp-solution}
s_\pm = \frac{2u_0(1-\xi) \pm \sqrt{s_0^2(1+\gamma) + u_0^2(2\xi - 1 + \gamma)^2}}{1+\gamma}.
\end{equation}

This relation distinctly manifests the lack of Galilean invariance in the graphene hydrodynamic equations. On the contrary, in 2D plasma of massive electrons, the velocities $s_{m\pm}$ are given by
\begin{equation}
s_{m\pm} = u_0 \pm s_0.
\end{equation} 

In graphene, the presence of steady flow modifies the properties of electron system, therefore, the plasma wave velocity $s_\pm$ depends on the flow velocity in a quite complicated manner described by Eq.~(\ref{Disp-solution}). Particularly, when the flow velocity is small ($\beta_0 \ll 1$), we obtain
\begin{equation}
s_\pm = 2 u_0 (1-\xi) \pm s_0.
\end{equation}
In a limit of strongly degenerate electrons ($\xi = 3/4$), the velocity of plasma waves is reduced to
\begin{equation}
\label{Plasma-velocity-simple}
s_\pm = \frac{1}{2} u_0 \pm  s_0.
\end{equation}

The factor $1/2$ formally originates due to the vanishing nonlinear convective term in the Euler equation. In case of large drift velocities $u_0$, the sign of convective term in Eq.~(\ref{Canonical-graphene}) can switch from positive to negative. This, in its turn, leads to a decrease in wave velocity $s_+$ with rising flow velocity. An increase in fictitious mass ${\cal M}$ also contributes to this process.

The wave velocities $s_\pm$ obtained from Eq.~(\ref{Disp-solution}) are plotted in Figs.~\ref{Spm} for different values of chemical potential, and for flow velocities ranging from zero up to $v_F$. An unusual 'halved' drag by the flow at small velocities turns to a decrease in wave velocity at large $u_0$. Formally, at $u_0 = v_F$ the velocities of both branches converge to $s_\pm=\pm v_F$. However, this case is of purely academic interest as such fast flows are unattainable owing to velocity saturation~\cite{Meric}.

\subsection{Excitation of electron plasma waves by direct current}
A special kind of plasma wave instability (Dyakonov-Shur instability) occurs in high-mobility field effect transistors under the condition of constant drain current~\cite{Dyakonov-Shur}. An amplification of plasma wave amplitude occurs after the reflection from the drain end. The corresponding increment $\omega_m ''$ (for 2D plasma of massive electrons) was shown to be governed by the ratio of forward and backward wave velocities: 
\begin{equation}
\omega_m '' = \frac{s_0^2-u_0^2}{2 L s_0}\ln\left[\frac{s_0 + u_0}{s_0 - u_0}\right].
\end{equation}

This plasma wave instability leads to radiation of electromagnetic waves due to oscillations of image charges in metal electrodes~\cite{Knap}. To find out whether such instability persists for the unusual electron dynamics in graphene, we solve Eqs.~(\ref{Eqn-of-continuity}) and (\ref{Eqn-of-motion}) with the boundary conditions
\begin{equation}
\label{Boundary_conditions}
\delta n|_{x=0}=0,\qquad \left.\left[ n_0 \delta u + u_0 \delta n\right]\right|_{x=L} = 0.
\end{equation}
The latter condition corresponds to a constant drain current. This can be realized either for transistors operating in the current saturation mode~\cite{Meric}, or with the help of an external circuit sustaining the constant current.
 
It is easy to show that the complex eigenfrequencies $\omega_n = \omega_n ' + i\omega_n ''$ of Eqs.~(\ref{Eqn-of-continuity}) and (\ref{Eqn-of-motion}) with boundary conditions (\ref{Boundary_conditions}) are
\begin{equation}
{\omega '_n} =\frac{\pi n}{2L}\frac{s_0^2 - u_0^2\left( 3-4\xi -\gamma  \right)}{\sqrt{s_0^2 \left( 1+\gamma  \right)+u_0^2{{\left( 2\xi -1+\gamma  \right)}^2}}},
\end{equation}
\begin{multline}
\label{Increment}
{\omega ''_{n=1}} = \frac{\omega '_{n=1}}{\pi}  \ln \left| \frac{s_+}{s_-}\right|\approx \\
\frac{2u_0(1-\xi)}{L} \frac{s_0^2 - u_0^2(3-4\xi-\gamma)}{s_0^2(1+\gamma)+u_0^2(1-2\xi-\gamma)^2}.
\end{multline}

The imaginary part of the complex frequency is positive, which means an amplification of the waves. We also see that for the existence of instability it does not matter whether the spectrum of electrons is parabolic or linear. The instability persists if only the velocities of forward and backward plasma waves are different. At the same time, compared with massive particles, the wave increment (\ref{Increment}) is smaller due to a smaller difference in wave velocities [$|s_+|-|s_-| \approx u_0$ for degenerate massless electrons instead of $2u_0$ for massive electrons]. As the flow velocity $u_0$ increases ($\beta_0 \gtrsim 1/2$), the wave increment begins to fall down because the velocity difference decreases.

\begin{figure}[ht]
\center{\includegraphics[width=1.0\linewidth]{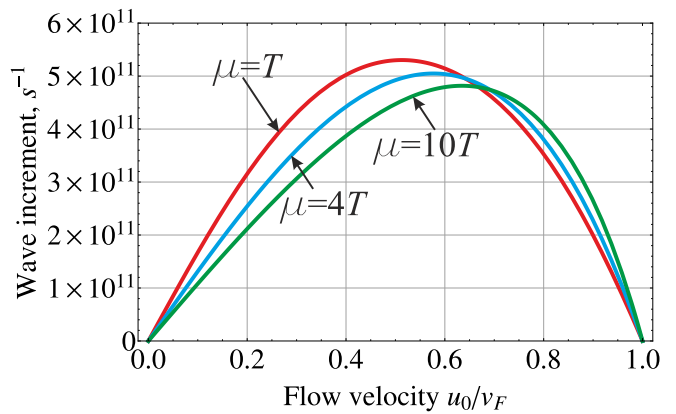} }
\caption{Dependence of wave increment $\omega_1''$ on flow velocity $u_0$ at different values of Fermi energy}
\label{Incr}
\end{figure}

The ultimate wave increment in such kind of instability is estimated as $v_F/(4L)$, which is attained at $u_0 \simeq v_F/2$ (see Fig.~\ref{Incr}). For the self-excitation to arise, it should exceed $(2\tau_p)^{-1}$, where $\tau_p$ is the momentum relaxation time. Considering the high-quality samples, where the scattering on acoustic phonons dominates, we can estimate $\tau_p^{-1} \approx 3\times 10^{11}$~s$^{-1}$ at room temperature~\cite{Vasko-Ryzhii}. The self-excitation turns out to be possible for channel lengths $L \lesssim 1.5$ $\mu$m. This optimistic anticipation could be hampered by the presence of impurity scattering, velocity saturation, and dependence of relaxation time on electron density. Nevertheless, even for shorter channels the hydrodynamic approach is valid and the self-excitation seems plausible.

\section{Discussion of the results}
We have derived hydrodynamic equations describing the transport of massless electrons in graphene. A linear energy spectrum of carriers should be taken into account from the very beginning of derivation. It cannot be introduced as a small correction to the parabolic dispersion (like $p^4$-terms in Si, Ge, A$_3$B$_5$~\cite{Nonparabolic-transport}). On the other hand, the hydrodynamic equations for ultrarelativistic plasmas cannot be also directly applied to graphene as the Fermi velocity is much smaller compared to that of light.

The dependence of particle density $n$ on drift velocity $u$ (Eq.~\ref{Eqn-of-state}) may look confusing. However, it is the immediate consequence of the particular choice of distribution function~(\ref{F_e}). At the same time, one can choose the distribution function in the form
\begin{equation}
\label{Conserving}
f ({\bf p}) = \left[1 + \exp \left(\frac{p v_F - {\bf p u}}{T (1-u^2/v_F^2)^{3/4}} -\frac{\mu}{T}\right)\right]^{-1}.
\end{equation}
This function turns collision integral to zero, reduces to the equilibrium Fermi function at $u \to 0$, and the corresponding particle density does not depend on $u$ (but the internal energy still does). It is easy to show that Euler and continuity equations derived with this function and written in terms of $n$ and $u$ coincide with Eqs.~(\ref{Euler}, \ref{Continuity}). The equation of state also holds its view. The boundary conditions for hydrodynamic equations are imposed on measurable quantities $n$ and $u$. Hence, the solutions of hydrodynamic equations do not depend on the choice between distribution functions (\ref{F_e}) and (\ref{Conserving}).

In the obtained hydrodynamic equations for electrons in graphene, the effect of linear spectrum is clearly visible. First, the drift velocity $u$ cannot overcome the Fermi velocity $v_F$. Secondly, a varying fictitious hydrodynamic mass ${\cal M} \approx {(\mu/v_F^2)}/\sqrt{1 - u^2/v_F^2}$ originates in the Euler equation. The obtained equations are neither Lorentz-, nor Galilean invaraint, which is directly revealed in the spectra of plasma waves in the presence of steady electron flow [Eq.~(\ref{Disp-solution})]. Our main conclusions concerning the spectra  can be verified experimentally using the techniques of plasmon nano-imaging~\cite{Plasmons-imaging} in gated graphene under applied bias. 

As we became aware recently, the spectra of plasma waves in the presence of steady flow and Dyakonov-Shur instability in graphene were analyzed in Ref.~\cite{Polini-hydro}. The form of Euler equation used was different from our Eq.~(\ref{Canonical-graphene}) even in the limit $u \ll v_F$, particularly, in Ref.~\cite{Polini-hydro}  the gradient term $u\partial_x u$ did not vanish for degenerate electron systems. This led to different expression for plasma wave velocities [$3u_0/4 \pm u_0$ instead of our Eq.~(\ref{Plasma-velocity-simple})], and to higher estimate of the plasma wave instability increment. One possible reason for discrepancy of equations lies in different expression for electron plasma pressure $P$, which is substantially velocity dependent [$P = \varepsilon/2 \propto\linebreak[4] \mu^3/(1-\beta^2)^{5/2}$, see Eqs. (\ref{Energy-density}) and (\ref{Stress-tensor})].

The set of problems which could be solved via nonlinear hydrodynamic equations is not restricted within plasma waves. It would be also interesting to study the effects of velocity saturation associated with the upper limit of drift velocity $u$ equal to $v_F$. Those effects could be pronounced in graphene samples on substrates with high optical phonon energy. If it is the case, the velocity saturation caused by emission of optical phonons~\cite{Meric} seems as irrelevant.

For rigorous simulation of emerging graphene-based devices for THz generation and detection~\cite{Graphene-photovoltaic} one can as well employ the derived nonlinear equations. In the case, however, an Euler equation for holes and electron-hole friction terms should be supplied~\cite{Our-hydrodynamics}. With large electron mobility and new hydrodynamic nonlinearities, graphene-based THz devices could outperform those based on conventional semiconductors.

\section{Conclusions}
The hydrodynamic equations governing the collective motion of massless electrons in graphene were derived. The validity of those equations is not restricted to small drift velocities. A variable fictitious mass depending on density and velocity arises in the hydrodynamic equations. It results in several nonlinear terms specific to graphene.

The possibility of soliton formation in electron plasma of the gated graphene was shown. The quasi-relativistic terms in the dynamic equations set an upper limit of the soliton amplitude and stabilize its shape.

The obtained hydrodynamic equations demonstrate the lack of Galilean and true Lorentz invariance. This non-invariance is pronouncedly revealed in the spectra of plasma waves in the presence of steady flow with velocity~$u_0$. The difference in velocities of forward and backward waves turns out to be $u_0$ instead of $2u_0$, expected for massive electrons in conventional semiconductors.

The possibility of plasma wave self-excitation in high-mobility graphene samples under certain boundary conditions (Dyakonov-Shur instability) was demonstrated. The increment of such instability in graphene is less than that in common semiconductors due to smaller difference in velocities of forward and backward waves. However, the high mobility of electrons in graphene allows plasma wave self-excitation for micron-length and shorter channels.

\section{Acknowledgement}
The work of D.S. and V.V. was supported by the Russian Ministry of Education and Science (grant 14.132.21.1687), Russian Foundation for Basic Research (grant 11-07-00464), and Russian Academy of Science. The work at RIEC was supported by the Japan Science and Technology Agency, CREST and by the Japan Society for Promotion of Science. The authors are grateful to S.O. Yurchenko for helpful discussions, and to prof. M. Polini for valuable comments.

\appendix
\section{Calculation of average values}
The statistical average values can be exactly calculated with the distribution function (\ref{F_e}) for linear energy spectrum. The particle density is given by
\begin{multline}
n = \frac{4}{{\left( 2\pi \hbar  \right)^2}}\int{\frac{pdpd\theta }{1+{e^{\frac{p\left( v_F - u\cos \theta  \right)-\mu}{T}}}}}=\\
\frac{T^2}{{{\left( \pi \hbar {v_F} \right)}^2}}\left\{ \int\limits_{0}^{\infty}{\frac{2\pi tdt}{1+{e^{t-\mu /T}}}} \right\}\left\{ \frac{1}{2\pi }\int\limits_{0}^{2\pi}{\frac{d\theta}{{{\left[ 1-\beta\cos \theta  \right]}^2}}} \right\}.
\end{multline}
The last term could be evaluated as $\left[1 - \beta^2\right]^{-3/2}$, while the remainder is nothing more but the particle density in the absence of flow $n_0$. Similarly, the expression for energy density reads
\begin{multline}
\label{Energy-density}
\varepsilon = \frac{4}{\left( 2\pi \hbar  \right)^2}\int{\frac{v_F p^2 dp d\theta }{1+e^{\frac{p\left( v_F - u\cos \theta  \right)-\mu}{T} }}}=\\
\frac{T^3}{\left( \pi \hbar {v_F} \right)^2} \left\{ \int\limits_{0}^{\infty}{\frac{2\pi t^2 dt}{1+{e^{t-\mu /T}}}} \right\} \left\{ \frac{1}{2\pi} \int\limits_{0}^{2\pi}{ \frac{d\theta}{{\left[ 1-\beta\cos \theta  \right]}^3}} \right\}.
\end{multline}
Here, the integral is evaluated as $\left[1 - \beta^2\right]^{-5/2}$, while the remainder is steady-state energy density $\varepsilon_0$.

The stress tensor is
\begin{multline}
\label{Stress-tensor}
\Pi_{xx} = \frac{4}{\left( 2\pi \hbar  \right)^2}\int{\frac{v_F p^2 \cos^2\theta dpd\theta }{1 + e^{\frac{p\left( v_F - u\cos \theta  \right)-\mu }{T}}}}=\\
\frac{T^3}{{{\left( \pi \hbar v_F \right)}^{2}}}\left\{ \int\limits_{0}^{\infty }{\frac{2\pi t^2 dt}{1+ e^{t-\mu /T}}} \right\}\left\{ \frac{1}{2\pi }\int\limits_{0}^{2\pi}{\frac{ \cos^2\theta d\theta }{{\left[ 1-\beta \cos \theta  \right]}^3}} \right\} = \\
\frac{\varepsilon}{2} (1+2\beta^2).
\end{multline}

The density of inverse energy $\langle \varepsilon^{-1} \rangle$ can be expressed in terms of elementary functions
\begin{equation}
\label{Inverse-energy}
\langle \varepsilon^{-1} \rangle = \frac{2T \ln\left[1+e^{\mu/T}\right]}{\pi \hbar^2 v_F^2 \sqrt{1-u^2/v_F^2}}.
\end{equation}

The following relations for the derivatives of average values are required to represent the Euler equation in the canonical form:
\begin{gather}
\label{Derivatives}
d n = \frac{1}{1 - \beta^2}\left( \left\langle  \varepsilon^{-1}  \right\rangle d \mu + 3 n \beta d \beta \right), \\
d  \varepsilon = \frac{1}{1 - \beta^2}\left( 2 n d \mu + 5  \varepsilon  \beta d \beta \right).
\end{gather}

\end{document}